\documentclass[final,3p,times,sort&compress]{elsarticle}

\usepackage{amsmath,amssymb}
\usepackage{bm}
\usepackage{graphicx}
\usepackage{color}
\usepackage[utf8]{inputenc}

\newcommand{\bhv}{\boldsymbol{\hat{\mathbf{v}}}}
\newcommand{\rmi}{\mathrm{i}}
\newcommand{\rmd}{\mathrm{d}}
\newcommand{\sfF}{\mathsf{F}}

\newcommand{\sfH}{\mathsf{H}}
\newcommand{\Hvac}{\mathsf{H}_\text{vac}}
\newcommand{\Hmat}{\mathsf{H}_\text{mat}}
\newcommand{\Hnu}{\mathsf{H}_{\nu\nu}}
\newcommand{\thetav}{\theta_\text{v}}
\newcommand{\Gf}{G_\text{F}}
\newcommand{\avg}[1]{\langle #1 \rangle}

\newcommand{\XFLAT}{\texttt{XFLAT}}
\newcommand{\NBeam}{\texttt{NBeam}}
\newcommand{\NBGroup}{\texttt{NBGroup}}

\begin{document}

\begin{frontmatter}

\title{Simulating collective neutrinos oscillations on the Intel Many
  Integrated Core (MIC) architecture} 

\author[ece]{Vahid Noormofidi}
\author[phy]{Susan R. Atlas}
\author[ece,phy]{Huaiyu Duan}
\ead{duan@unm.edu}

\address[ece]{Department of Electrical and Computer Engineering,
  University of New Mexico, Albuquerque, NM 87131, USA}
\address[phy]{Department of Physics and Astronomy, University of New
  Mexico, Albuquerque, NM 87131, USA}




\begin{abstract}
We evaluate the second-generation Intel Xeon Phi coprocessor
based on the Intel
Many Integrated Core (MIC) architecture, aka
the Knights Landing or KNL, for
simulating neutrino oscillations in (core-collapse) supernovae. For
this purpose we have
developed a numerical code \XFLAT\ which is optimized
for the MIC architecture and which can run on both the homogeneous HPC platform
with CPUs or Xeon Phis only and the hybrid platform with both CPUs and
Xeon Phis. To efficiently utilize 
the SIMD (vector) units of the MIC architecture we implemented a
design of Structure of Array (SoA) in the low-level module of the
code. 
We benchmarked the code on the NERSC Cori supercomputer which is equipped with dual 68-core
7250 Xeon Phis.  
We find that compare to the first generation of the Xeon Phi (Knights Corner a.k.a KNC) the performance improves by many folds. 
Some of the problems that we encountered in this work may be solved with
the advent of the new supernova model for neutrino oscillations and
the next-generation Xeon Phi.  
\end{abstract}


\end{frontmatter}


\section{Introduction}

At the end of its life, the core of a massive star collapses under its
own gravity to a neutron star or black hole, and 
the rest of the star explodes as a (core-collapse) supernova (see,
e.g., Ref.~\cite{Woosley:2005yv} for a review). Supernovae%
\footnote{The phrase ``supernova'' in this article refers exclusively
  to the core-collapse supernova. There are also type Ia supernovae which
  do not emit substantial amount of neutrinos.}
are essential to the chemical evolution of the universe. They enrich the
interstellar medium with heavy elements which are synthesized during the stellar
evolution and explosion. It is in the ashes of numerous supernovae
that our solar system was born.

A gigantic amount of the gravitational binding energy
($\sim3\times10^{46}\,$J) of the stellar core is released during the
collapse. Because the
nascent neutron star formed at the center of the supernova is so dense 
($\gtrsim 10^{14}\,\text{g}\,\text{cm}^{-3}$) that almost all
particles including photons are trapped within it. The vast majority
of the energy ($\sim99\%$) is carried away by
a group of nimble particles called neutrinos ($\nu$) which have no electric
charge and interact very weakly with ordinary matter.
Neutrinos have three species or ``flavors'': electron ($e$),
muon ($\mu$) and 
tau ($\tau$) flavors (which are named after their weak-interaction partners).
One of the recent major 
breakthroughs in particle physics is the
discovery that neutrinos of different flavors can transform or
``oscillate'' into each 
other while they propagate in space, a quantum phenomenon known as neutrino
oscillations (see, e.g., Ref.~\cite{Agashe:2014kda} for a review). 
The knowledge of exactly how neutrinos oscillate in the supernova environment
is an important and yet missing piece in the puzzle of
how supernovae 
explode and what elements are produced in such environment. This
knowledge is also essential to the correct interpretation of the 
neutrino signals from future galactic supernovae
\cite{Mirizzi:2015eza}.

Although the supernova envelope is essentially transparent to
neutrinos, the refractive indices of the neutrinos in matter are flavor
dependent and are also different from those in vacuum. In addition, there
are $\sim10^{58}$ neutrinos emitted from the neutron star (of radius
$10-60\,\text{km}$) in just tens of seconds. These neutrinos form a
dense neutrino medium surrounding the neutron star which can potentially
oscillate collectively as a whole (see, e.g., Ref.~\cite{Duan:2010bg}
for a review). The full information of the flavor oscillations of this
neutrino medium can be achieved by solving a seven-dimensional quantum
kinetic equation (with one 
temporal dimension, three spatial dimensions and three momentum
dimensions) which is such a 
challenging task
that this equation itself was obtained only recently  \cite{Vlasenko:2013fja}.
Almost all the existing work on supernova neutrino oscillations is
based on a much simplified model known as the (neutrino) bulb model with
only one spatial dimension and two momentum dimensions
\cite{Duan:2006an}. Neutrino oscillations in the bulb model are
described by millions of 
coupled nonlinear equations which can be solved numerically 
on a computer cluster. Because each additional dimension can easily
increase the computation load by a few magnitudes, solving
the full seven-dimensional model demands
computation resources which are available only on
next-generation supercomputers.

Some of the next-generation supercomputers are expected to employ
the Intel Many Integrated Core (MIC) architecture.
The MIC architecture can significantly boost 
parallel computation 
with its many simple x86-like cores and wide (512-byte) SIMD
(i.e.\ same instruction, multiple data) vector units. The first
generation Intel Xeon Phi coprocessor based on MIC known as the
Knights Corners or KNC can achieve a peak
performance of more than 1 teraFLOPS for double precision calculations.
Because MIC supports both standard programming
languages (C, C++ and Fortran) and standard parallel programming
interfaces (such as OpenMP and MPI), it is possible to maintain the same
code base for both MIC and CPU based supercomputers or even the hybrid
supercomputers with both CPUs and MIC-based coprocessors.
 
In this work we evaluate the MIC for simulating neutrino oscillations in
supernovae. In Section~\ref{sec:model} we describe the physics of the bulb model
used for the simulation. In Section~\ref{sec:implementation} we
outline the numerical implementation of the bulb model and highlight
the major optimization of \XFLAT\ for the MIC architecture. In
Section~\ref{sec:performance} we present the performance benchmarks of
\XFLAT\ on the Stampede 
supercomputer at the Texas Advanced Computing Center (TACC) with
various configurations.
In Section~\ref{sec:discussion} we discuss the advantages and issues
with the (current-generation) Xeon Phi and conclude this work.

\section{Neutrino oscillations in the extended bulb model%
\label{sec:model}}

\begin{figure}
\centering
\includegraphics[width=.5\textwidth]{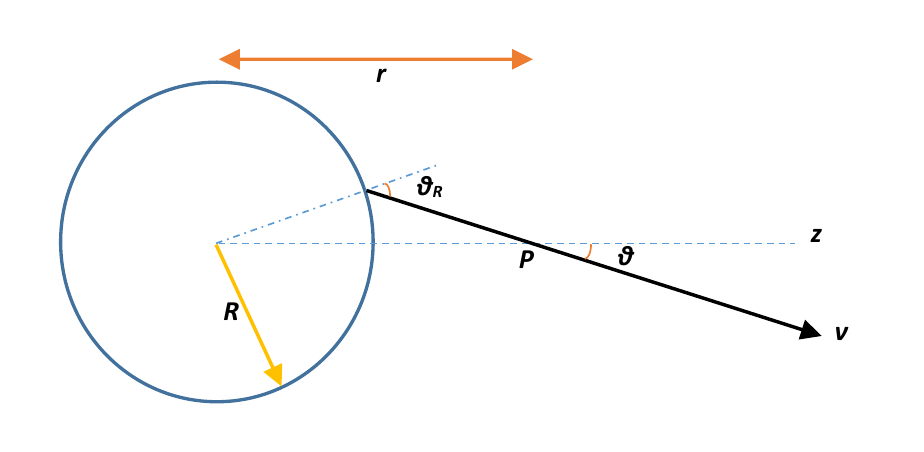}
\caption{Geometry of the (neutrino) bulb model. Because of the spherical
  symmetry around the center of the supernova and the axial symmetry
  around the radial ($z$) direction, a neutrino beam is uniquely
  determined by its emission angle $\vartheta_R$ on the surface of the
  neutron star (of radius $R$). The polar angle $\vartheta(r)$ which the
  neutrino beam makes with the $z$ direction at point $P$ depends on
  the distance $r$ of $P$ from the center of the supernova.
  The azimuthal angle $\varphi$ around the $z$ axis (not shown) is also
  needed to describe the direction of the neutrino beam in the
  extended bulb model which does not have the axial symmetry around
  the $z$ axis.}
\label{fig:bulb-model}
\end{figure}

In the bulb model it is assumed that both the neutrino
emission and the supernova environment are stationary because
the evolution timescale
of the supernova is much longer than that of neutrino
oscillations. It is further assumed that the supernova is spherically
symmetric around its center so that only radius $r$ or the distance
from the center of the supernova is needed to describe the spatial
coordinate the neutrino. All the neutrinos are emitted from the
surface of the spherical neutron star of radius $R$ and stream through
the supernova envelope with the speed of light $c$ without hindering.
The neutrino emission in the bulb model is assumed to be
axially symmetric around the radial direction. 
Therefore, the propagation direction  of a neutrino beam 
is uniquely determined by its emission angle
$\vartheta_R=\vartheta(R)$ on the surface of the neutron star or,
equivalent, by the polar angle $\vartheta(r)$ between the propagation
direction of the neutrino and the radial direction (see
Fig.~\ref{fig:bulb-model}).

The flavor
quantum state of a neutrino or antineutrino (i.e.\ the antiparticle of
the neutrino) in the bulb model is described by
a multi-component complex wavefunction
\begin{align}
\psi_{\nu_\alpha}(E,\bhv;r) = \begin{pmatrix}
a \\ b
\end{pmatrix}
\qquad\text{or}\qquad
\psi_{\bar\nu_\alpha}(E,\bhv;r) = \begin{pmatrix}
\bar{a} \\ \bar{b}
\end{pmatrix},
\label{eq:psi}
\end{align}
where $\alpha$ is the initial flavor of the neutrino at radius $R$,
$E$ is the energy of the neutrino, and $\bhv$ is the unit vector
along the propagation direction of the neutrino. As in
Ref.~\cite{Duan:2006an} we assume a flavor mixing between two
flavors $e$ and $x$ only, where $\nu_x$ is a linear combination of
$\nu_\mu$ and $\nu_\tau$. Generalization to the three-flavor
mixing can also be done \cite{Duan:2007sh,Cherry:2010yc}. In
Eq.~\eqref{eq:psi} $a$ and $b$ are the amplitudes of the neutrino to
be in the $e$ and $x$ flavors, respectively, and, therefore, $|a|^2$
and $|b|^2$ give the probabilities for the neutrino to be detected in
the corresponding flavors.

The flavor content of the neutrino medium can solved from the
Schr\"odinger equation 
\begin{subequations}
\label{eq:eom}
\begin{align}
\rmi \cos\vartheta\, \partial_r \psi_{\nu_\alpha}(E,\bhv;r)
&= \sfH\, \psi_{\nu_\alpha}(E,\bhv;r)
= [\Hvac(E) + \Hmat(r) + \Hnu(\bhv;r)]\, \psi_{\nu_\alpha}(E,\bhv;r), \\
\rmi \cos\vartheta\, \partial_r \psi_{\bar\nu_\alpha}(E,\bhv;r)
&= \bar\sfH\, \psi_{\bar\nu_\alpha}(E,\bhv;r)
= [\Hvac(E) - \Hmat(r) - \Hnu^*(\bhv;r)]\, \psi_{\bar\nu_\alpha}(E,\bhv;r), 
\end{align}
\end{subequations}
where $\partial_r$ is the partial differentiation with respect to
$r$, $\Hvac$ is the standard vacuum Hamiltonian, and $\Hmat$ and
$\Hnu$ are the matter and neutrino potentials, respectively. 
Here we have adopted the natural units with both the (reduced) Planck constant
$\hbar$ and the speed of light $c$ equal to 1.
In the two-flavor mixing scenario the vacuum Hamiltonian
\begin{align}
\Hvac(E) = \frac{\Delta m^2}{4 E}\, \begin{pmatrix}
-\cos 2\thetav & \sin2\thetav \\
\sin2\thetav & \cos2\thetav
\end{pmatrix}
\end{align}
depends on the neutrino energy $E$ and two key parameters of neutrino
mixing: the mass-squared difference $\Delta m^2$ 
and the  vacuum mixing angle $\thetav$ \cite{Agashe:2014kda}.
The matter potential is
\begin{align}
\Hmat(r) = \sqrt2 \Gf n_e(r)\, \begin{pmatrix}
1 & 0 \\
0 & 0
\end{pmatrix},
\end{align}
where $\Gf\approx 1.166\times10^{-5}\,\text{GeV}^{-2}$ is the Fermi
coupling constant for the weak interaction, and $n_e(r)$ is the electron
number density at radius $r$. The difficulty of solving
Eq.~\eqref{eq:eom} stems from the neutrino potential
\begin{align}
\Hnu(\bhv;r) = \frac{\sqrt2\Gf}{2\pi R^2}
\sum_{\alpha'} \int_0^1\rmd(\cos\vartheta')\int_0^\infty\rmd E'\,
(1-\bhv\cdot\bhv')\, \sfF_{\alpha'}(E';\bhv';r), 
\label{eq:Hnu}
\end{align}
which couples the flavor evolution of all the neutrinos and
antineutrinos. In the above equation
\begin{align}
1-\bhv\cdot\bhv' = 1-\cos\vartheta\cos\vartheta',
\end{align}
and
\begin{align}
\sfF_{\alpha}(E,\bhv;r)
= \frac{L_{\nu_\alpha}}{\avg{E_{\nu_\alpha}}}\, 
f_{\nu_\alpha}(E)\,
 [\psi_{\nu_\alpha}(E,\bhv;r)\, \psi_{\nu_\alpha}^\dagger(E,\bhv;r)]
-\frac{L_{\bar\nu_\alpha}}{\avg{E_{\bar\nu_\alpha}}}\, 
f_{\bar\nu_\alpha}(E) \,
[\psi_{\bar\nu_\alpha}(E,\bhv;r)\, \psi_{\bar\nu_\alpha}^\dagger(E,\bhv;r)]^*,
\end{align}
where $L$, $\avg{E}$ and $f(E)$ are the energy luminosities, average
energies and normalized energy distribution functions of the
correspond neutrinos and antineutrinos on the surface of the neutron star.

It has been shown recently that the axial symmetry
of the neutrino flavor wavefunction around the radial direction can be broken
spontaneously by collective neutrino oscillations
\cite{Raffelt:2013rqa,Mirizzi:2013rla}. In this case it seems natural
to generalize the bulb model to the extended bulb model in which $\bhv$ is
determined by both the 
azimuthal angle $\varphi$ around the radial direction and
the polar angle $\vartheta$. Correspondingly, one should make the
replacements
\begin{align}
1-\bhv\cdot\bhv'
&\longrightarrow
1-[\cos\vartheta\cos\vartheta'
+\sin\vartheta\sin\vartheta'
(\cos\varphi\cos\varphi'+\sin\varphi\sin\varphi')],\\
\int_0^1\rmd(\cos\vartheta')
&\longrightarrow
\frac{1}{2\pi}\int_0^{2\pi}\rmd\varphi' \int_0^1\rmd(\cos\vartheta')
\label{eq:int-v}
\end{align}
in Eq.~\eqref{eq:Hnu}.
However, we note that the extended bulb model is not a
self-consistent model because it is not possible to maintain the (spatial)
spherical symmetry about the center of the neutron star without the
(directional) axial symmetry around the radial
direction. Nevertheless, because of 
the lack of a better model, we will use
the extended bulb in our evaluation of the MIC architecture.

We note that the neutrino potential $\Hnu(\bhv;r)$ in
Eq.~\eqref{eq:Hnu} is the same for all the neutrinos with the same
propagation direction $\bhv$ and does not depend on the energy $E$ of the
neutrino. Further, in the extended bulb model the same few weighted
integrals are needed in computing $\Hnu(\bhv;r)$ for all neutrino beams:
\begin{subequations}
\label{eq:F}
\begin{align}
\Phi_1(r) &=
         \sum_\alpha\int_0^1\rmd(\cos\vartheta)
\int_0^{2\pi}\rmd\varphi
\int_0^\infty\rmd E\,
\sfF_\alpha(E, \bhv; r), \\
\Phi_\mathrm{c}(r) &=
         \sum_\alpha\int_0^1\rmd(\cos\vartheta) \,\cos\vartheta
\int_0^{2\pi}\rmd\varphi
\int_0^\infty\rmd E\,
\sfF_\alpha(E, \bhv; r), \\
\Phi_\mathrm{sc}(r) &=
         \sum_\alpha \int_0^1\rmd(\cos\vartheta) \,\sin\vartheta
\int_0^{2\pi}\rmd\varphi\,\cos\varphi
\int_0^\infty\rmd E\,
\sfF_\alpha(E, \bhv; r), \\
\Phi_\mathrm{ss}(r) &=
         \sum_\alpha \int_0^1\rmd(\cos\vartheta) \,\sin\vartheta
\int_0^{2\pi}\rmd\varphi\,\sin\varphi
\int_0^\infty\rmd E\,
\sfF_\alpha(E, \bhv; r).
\end{align}
\end{subequations}
This observation simplifies the numerical implementation greatly.

\section{Numerical implementation%
\label{sec:implementation}}

We developed a C++ code XFLAT largely based on the algorithm explained
in Ref.~\cite{Duan:2008eb}. Here we highlight the modifications and optimization
for the MIC architecture \cite{NoormofidiDissertation}. 

\subsection{Utilization of Xeon Phi}

The first generation Xeon Phi KNC is in the form of coprocessor or
accelerator packaged as a PCIe card. The second generation Xeon Phi is
or will be in both forms of processor and coprocessor. Even though KNC
works as an add-on accelerator, it supports the native mode in which
it runs its own OS and can function as an independent MPI
node. In implementing XFLAT we decided to take advantage of the
similarities of the x86 and MIC architectures and created a single code
base which can be compiled and run on both architectures. When MPI is
enabled, a separate MPI task is run on each CPU and Xeon Phi
accelerator which we now loosely refer to a ``computing unit''. In
each MPI task OpenMP threads employed to exploit the multi-processing
power of each computing unit.

\subsection{Discretization and data structure}

In \XFLAT\ 
a group or ``beam'' of neutrinos with the same propagation direction
$\bhv$ and initial particle identity (i.e.\ $\nu_e$, $\nu_x$,
$\bar\nu_e$ or $\bar\nu_x$) is described by an \NBeam\ object.%
\footnote{We use the phrase ``object'' in a broad sense which can
  refer to either a class or an instance of the class in C++.}
For the two-flavor mixing scenario an \NBeam\ object consists of 4
floating point arrays, \texttt{ar[E\_CNT]}, \texttt{ai[E\_CNT]},
\texttt{br[E\_CNT]} and \texttt{bi[E\_CNT]}, which are the real and
imaginary components of the neutrino wavefunctions of the neutrinos in
the beam [see Eq.~\eqref{eq:psi}]. For simplicity we have discretized
the continuous energy 
distributions or energy spectra of the neutrino into energy bins of
equal energy intervals, and \texttt{E\_CNT} is the total number of
energy bins.

The flavor quantum state 
$\{\psi_{\nu_\alpha}(E,\bhv;r), \psi_{\bar\nu_\alpha}(E,\bhv;r)\}$ of
the neutrino medium at radius $r$ is represented by a three dimensional
object array \texttt{NBeam[THETA\_CNT][PHI\_CNT][PARTICLE\_CNT]}, where
\texttt{THETA\_CNT}, \texttt{PHI\_CNT} and \texttt{PARTICLE\_CNT} are
the numbers of polar angle ($\vartheta$) bins, azimuthal angle
($\varphi$) bins and initial particle identities, respectively.
The continuous neutrino flux distribution in the polar
angle $\vartheta$ and azimuthal angle $\varphi$ 
is discretized as equal-sized bins of
$u=\sin^2\vartheta_R$ and $\varphi$ within ranges
$u\in[0,1]$ and $\varphi\in[0,2\pi)$, respectively.
We chose this discretization scheme because, if the overall neutrino
flux is uniform in terms of the differential solid angle
$\rmd(\cos\vartheta)\,\rmd\varphi$, it will also be (almost) uniform
in terms of $\rmd u\,\rmd\varphi$ at $r\gg R$. 

In \XFLAT\ the first dimension (in polar angle
$\vartheta$) of the \texttt{NBeam} array is actually divided into
several sections each of which is located in the memory of a separate compute
device (i.e., a CPU or Xeon Phi accelerator). The number of polar angle bins,
\texttt{THETA\_CNT\_LOCAL}, on each device depends on its relative computing
capability.

\subsection{SoA data structure and SIMD parallelization}

The \NBeam\ object in \XFLAT uses a Structure-of-Arrays (SoA)
arrangement. In contrast Ref.~\cite{Duan:2008eb} defines the \NBeam\ object
as an Array of Structures (AoS). In the AoS arrangement
an \NBeam\ object consists of an array \texttt{PSI[E\_CNT]}.
Each element of this array is an object \texttt{PSI} which represents
the wavefunction of a single neutrino and consists of 4 floating point numbers,
\texttt{ar}, \texttt{ai}, \texttt{br} and \texttt{bi}.

The Intel Xeon Phi accelerator is capable of performing same
instructions on multiple data (SIMD) 
simultaneously. The SoA implementation of the \NBeam object in \XFLAT\ 
facilitates the SIMD parallelization provided by the MIC
architecture. 

\subsection{Numerical algorithm and parallelization}

\XFLAT\ solves Eq.~\eqref{eq:eom} as an initial condition problem with
initial conditions 
\begin{align}
\psi_{\nu_e}(E,\bhv;r=R)
&= \psi_{\bar\nu_e}(E,\bhv;r=R)
=\begin{pmatrix}
1 \\ 0
\end{pmatrix}, &
\psi_{\nu_x}(E,\bhv;r=R)
&=\psi_{\bar\nu_x}(E,\bhv;r=R)
=\begin{pmatrix}
0 \\ 1
\end{pmatrix}. 
\end{align}
The basic algorithm that evolves the neutrino wave function over a
small radial step $\delta r$ is \cite{Duan:2006an}
\begin{subequations}
\label{eq:U}
\begin{align}
\psi_\nu(E,\bhv;r+\delta r)
&=\exp(-\rmi\sfH\, \delta l)\, \psi_\nu(E,\bhv; r)\\
&=\frac{1}{\lambda}\begin{pmatrix}
\lambda \cos(\lambda\delta l) -\rmi h_{11}\sin(\lambda \delta l) &
-\rmi h_{12} \sin(\lambda \delta l) \\
-\rmi h_{12}^* \sin(\lambda \delta l) &
\lambda \cos(\lambda\delta l) +\rmi h_{11}\sin(\lambda \delta l)
\end{pmatrix}
\psi_\nu(E,\bhv; r),
\end{align}
\end{subequations}
where $\delta l=\delta r/\cos\vartheta$, $h_{11}$ and $h_{12}$ are
the diagonal and off-diagonal elements of 
the total Hamiltonian $\sfH$, respectively, and
$\lambda = \sqrt{h_{11}^2+|h_{12}|^2}$.
On top of this algorithm \XFLAT\ uses a modified mid-point integration
solver with adaptive step-size control \cite{Duan:2008eb}.

\begin{figure}
\centering
\includegraphics[width=0.9\textwidth]{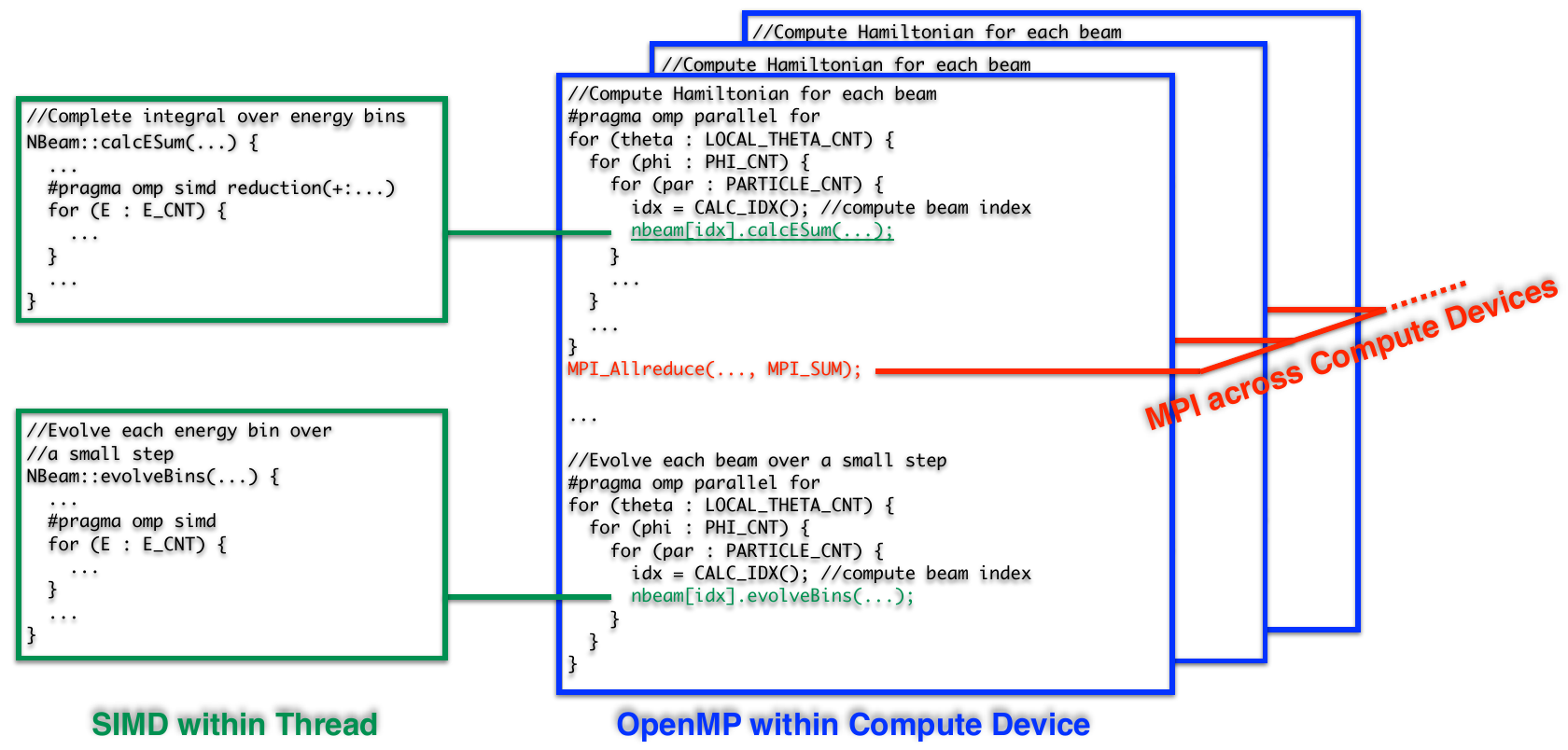}
\caption{The \texttt{evolveBins} method of the \NBeam{} object which evolves the
  wavefunctions in the neutrino beam for a small radial step by using
  the procedure described in Eq.~\eqref{eq:U}. The \texttt{omp}
  \texttt{simd} pragma instructs the compiler to generate a vectorized
  code which 
  utilizes the SIMD units on MIC.}
\label{fig:parallel}
\end{figure}

\XFLAT\ implements a parallel version of the above algorithm which
utilizes all three levels of parallelism
provided by the MIC architecture, i.e., SIMD within an thread, OpenMP
threads within a compute device and MPI among different compute
devices (see Fig.~\ref{fig:parallel}). 

In computing the neutrino potential
$\Hnu$ [see Eqs.~\eqref{eq:Hnu} and \eqref{eq:int-v}]
the innermost integral over the neutrino energy in
Eq.~\eqref{eq:F} is performed by a \texttt{for} loop with the OpenMP
SIMD directive.

is computed by the summation or integration of the
contributions from all  
neutrino beams [see Eqs.~\eqref{eq:Hnu} and \eqref{eq:int-v}]. 
This summation is carried out by using all three levels of parallelism
provided by the MIC architecture. At the top level the whole range of
the polar angle $\vartheta$ is broken down into several sections, and
the  
At the
lowest level the summation over neutrino energy bins is performed
inside each OpenMP thread by using the OpenMP SIMD directive. This
corresponds to the innermost integral over the neutrino energy in
\eqref{eq:F}. 
in
Eqs.~\eqref{eq:Hnu}, \eqref{eq:F} and \eqref{eq:U}. 

However, this Array-of-Structure (AoS) design does not work well
with the SIMD units on MIC or the modern CPUs (see, e.g.,
\cite{JeffersReinders201303}). Instead, we adopt a
Structure-of-Array (SoA) design and define four arrays in each
\NBeam{} object: \texttt{ar}$[E]$, \texttt{ai}$[E]$, \texttt{br}$[E]$ and
\texttt{bi}$[E]$, which represent the real and imaginary parts of the
upper and lower components of the wavefunction in
Eq.~\eqref{eq:psi}. The \texttt{omp} \texttt{simd} pragma is then used
to instruct the compiler to vectorize the code so that
the procedure in Eq.~\eqref{eq:U} can be applied to multiple neutrino
wavefunctions 
(8 for KNC and 4 for the E5-2869 Xeon CPU) simultaneously 
through the SIMD units (see Fig.~\ref{fig:simd}).

\begin{figure}
\centering
\includegraphics[width=.5\textwidth]{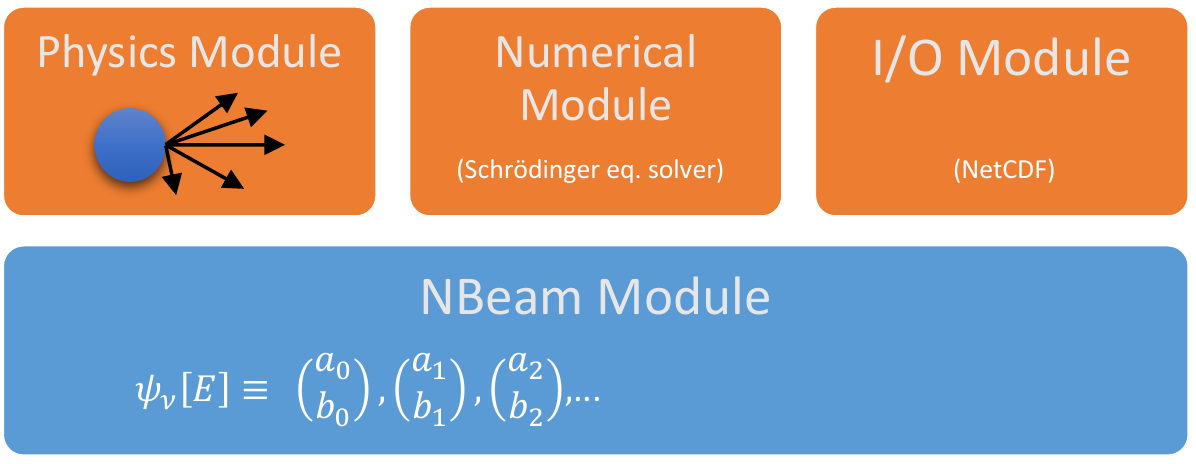}
\caption{The dual-layer design of XFLAT. The \NBeam{} module on the
  lower-level implements
  an array of the wavefunctions of the neutrinos with the same initial
  flavor and propagation direction. It adopts the SoA (Structure of
  Array) data structure to utilize 
  efficiently the SIMD units on MIC. The upper-level modules implement the
  physics and geometry of the supernova model, the numerical algorithm
  for solving the Schr\"odinger equation and the data I/O,
  respectively. They adopt the AoS (Array
  of Structure) data structure and are designed to be flexible.}
\label{fig:modules}
\end{figure}

XFLAT adopts a dual-layer modular design (see Fig.~\ref{fig:modules}). The
lower-level module defines the \NBeam{} object. On top this module
are three modules which operate closely with each other and
carry out different tasks:
\begin{itemize}
\item The Physics Module describes the physics and geometry of the
  supernova model and
  provides the methods for computing the matter potential $\Hmat$ and
  the neutrino potential $\Hnu$.

\item The Numerical Module implements a modified midpoint method
  described in Ref.~\cite{Duan:2008eb} to solve the differential
  equation~\eqref{eq:eom}.

\item The I/O Module performs data input and output by using the
 netCDF library \cite{netCDF}.
\end{itemize}
Unlike the NBeam Module which is optimized for the SIMD units by with
the SoA data structure, the three upper-level modules use the AoS
design to take advantage of the Object-Oriented Programming (OOP)
paradigm of C++. For example,  the Physics
Module defines the \NBGroup{} object which include all the neutrino
beams and is essentially an array of \NBeam{} objects.

As in Ref.~\cite{Duan:2008eb} each of the four modules
of XFLAT is carefully crafted to be independent of the internal details of
other modules. For example, we have implemented the Physics Modules
for both the bulb model and the extended bulb model. One can use XFLAT
to carry
out the calculations for either model by choosing the appropriate
Physics Module without modifying the rest of the code.
 
XFLAT uses all the three levels of parallelism provided by MIC. At the
top level the MPI is used to divide the full \NBeam{} array of the
\NBGroup{} object into subarrays and assign them to different MPI
processes or ranks on the available compute devices (CPUs and/or Xeon Phis).
XFLAT utilizes the native mode of the Xeon Phi coprocessor such that
each Xeon Phi runs an independent MPI process and is treated as
an ``ordinary'' CPU compute node except with more cores and wider SIMD
units.  
At the middle level the operations on the subarray of \NBeam{}
objects is carried out in parallel on the
many cores (and hyperthreads) on the CPU or Xeon Phi
via OpenMP. At the bottom level the operations on 
the individual neutrino wavefunctions inside each \NBeam{} object are
performed in parallel on the SIMD units using the \texttt{omp}
\texttt{simd} pragma. For example, the
integration over energy $E$ in Eq.~\eqref{eq:F} is carried out within each
thread using the SIMD units with the \texttt{omp} \texttt{simd}
\texttt{reduction} pragma, and the integration over angles $\vartheta$
and $\varphi$ is performed first inside each MPI process using OpenMP
with the 
\texttt{omp} \texttt{parallel}  \texttt{for} \texttt{reduction}
pragma and then across the MPI processes with the \texttt{MPI\_Allreduce}
function call.

\begin{figure}[h!]
\centering
\includegraphics[width=.49\textwidth]{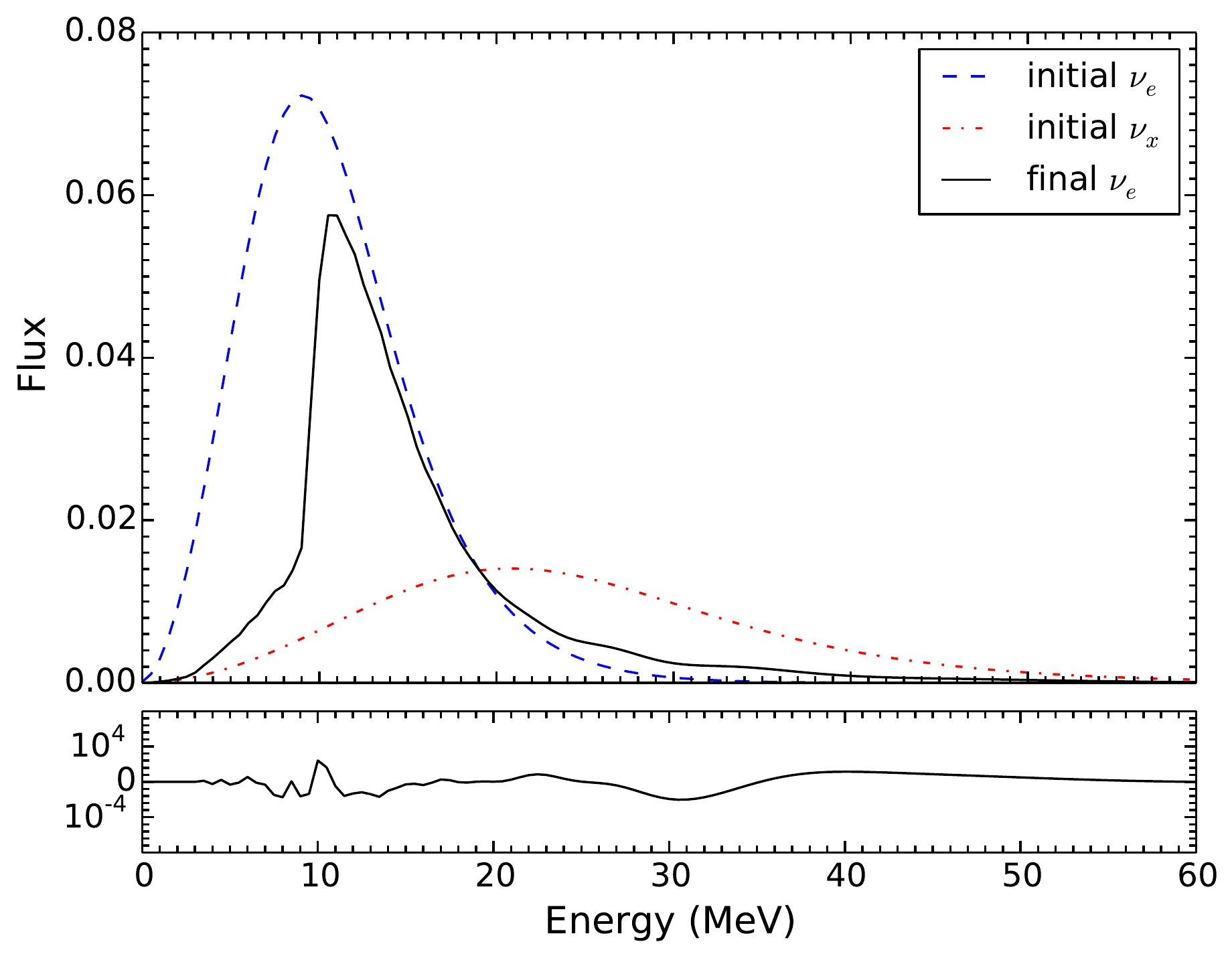}
\includegraphics[width=.49\textwidth]{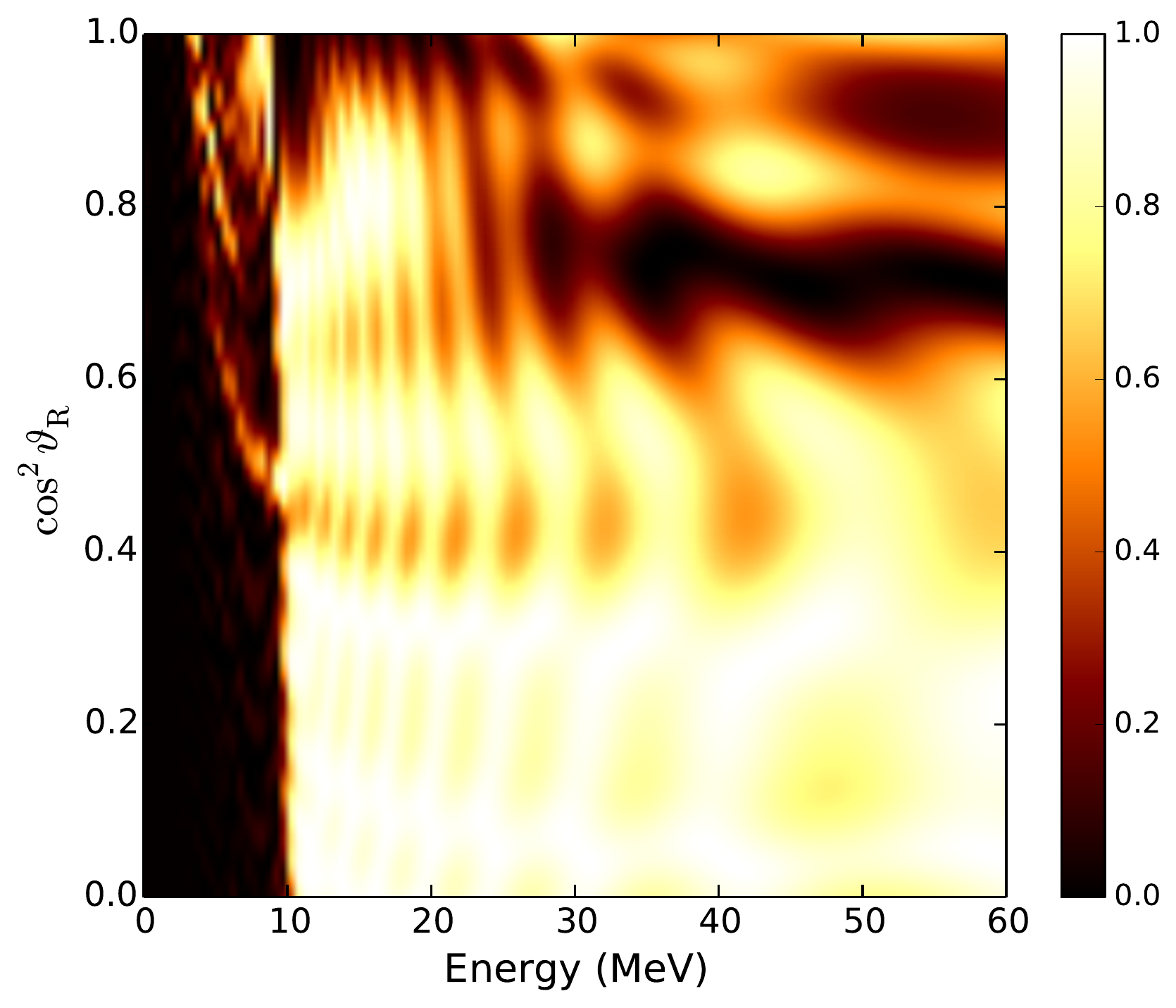}
\caption{Top left: The initial number fluxes of the neutrino in the
  electron ($\nu_e$, dashed curve) and the other flavor
  ($\nu_x$, dotted curve) in some arbitrary units on the surface of the
  neutron star and the $\nu_e$ flux at $r=250\,\text{km}$ as
  functions of neutrino energy $E$ in the bulb
  model as computed by XFLAT. Bottom left: The absolute difference
  between the $\nu_e$ fluxes at $r=250\,\text{km}$ between the XFLAT
  result and that of the ``SS spectra'' case in Ref.~\cite{Duan:2014mfa}.
  Right: Neutrino survival probability $P_{ee}=|a_e|^2$
  (represented by the color scale) at $r=250\,\text{km}$ in the bulb
  model computed by XFLAT as
  a function of the neutrino energy $E$ and emission angle
  $\vartheta_R$ on the surface of the neutron star.}
\label{fig:validation}
\end{figure}

We validate XFLAT by comparing its results for the bulb model and the
extended bulb model for the scenarios where the axial symmetry around the radial
direction is not broken.
We also compare its results for the bulb model with those in the
literature  \cite{NoormofidiDissertation}.
In Fig.~\ref{fig:validation} we show the results of the bulb model
computed using XFLAT using the same neutrino mixing parameters and the
so-called ``single-split spectra'' in
Ref.~\cite{Duan:2014mfa}. The results produced by XFLAT agree very
well with those in Ref.~\cite{Duan:2014mfa} which are based on a
numerical code developed 
independently at the Northwestern University \cite{ShalgarDissertation}.

\section{Performance analysis%
\label{sec:performance}}

We analyze the performance of XFLAT on the KNL nodes of the Cori supercomputer at
NERSC (see
Table~\ref{tab:nersc} for its hardware specifications). Double
precision was used exclusively in all floating point
calculations. For all the benchmarks reported here XFLAT was compiled
with the Intel C++ 
compiler (v18.0.3) with flags \texttt{-O3} \texttt{-openmp}
\texttt{-xMIC-AVX512}. One XFLAT 
process was run on each KNL. Each process was run with 272 OpenMP threads to fully utilize its hyperthreading
capability. For MPI-enabled benchmarks the Intel MPI library
2018 was used. Unless stated otherwise, all the
calculations are for a ``standard problem''

\begin{table}[h]
\centering
\caption{The hardware specifications of the
  compute nodes of the Cori supercomputer \cite{Cori}.}
\renewcommand{\arraystretch}{1.2}
\begin{tabular}[t]{l l}
\hline\hline
Component & Specifications \\ \hline
CPU & $1\times$ Knights Landing 7250 (68 cores, 272 threads) @1.4GHz\\
Memory  & 96GB ($6\times$ 16GB) DDR4 @2400MHz\\ 
MCDRAM & 16GB on-package, high-bandwidth memory\\
\hline\hline
\end{tabular}
\label{tab:cori}
\end{table}

\subsection{Single-node performance}

\begin{figure}[t]
\centering
\includegraphics[width=.49\columnwidth]{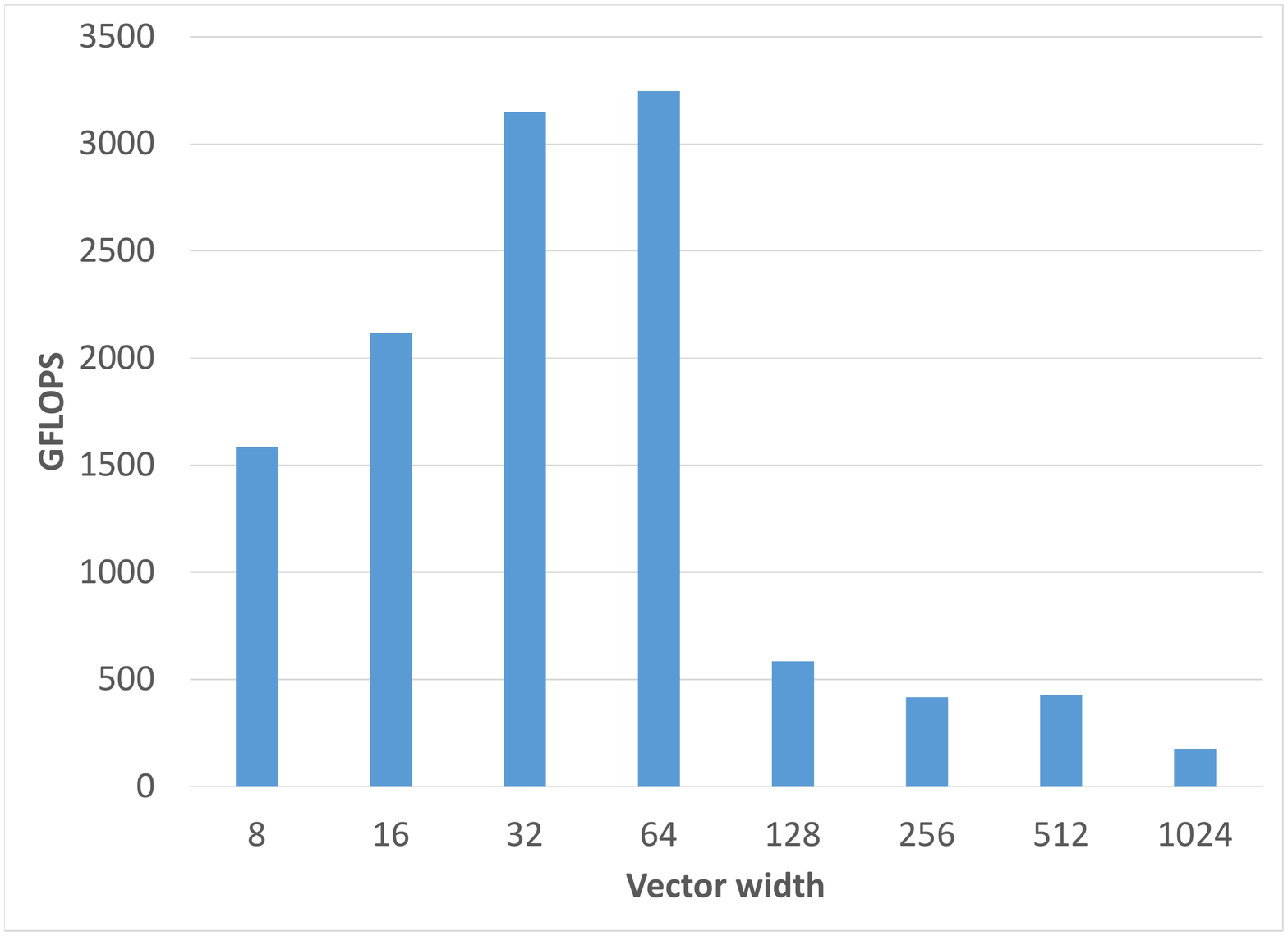}
\includegraphics[width=.49\columnwidth]{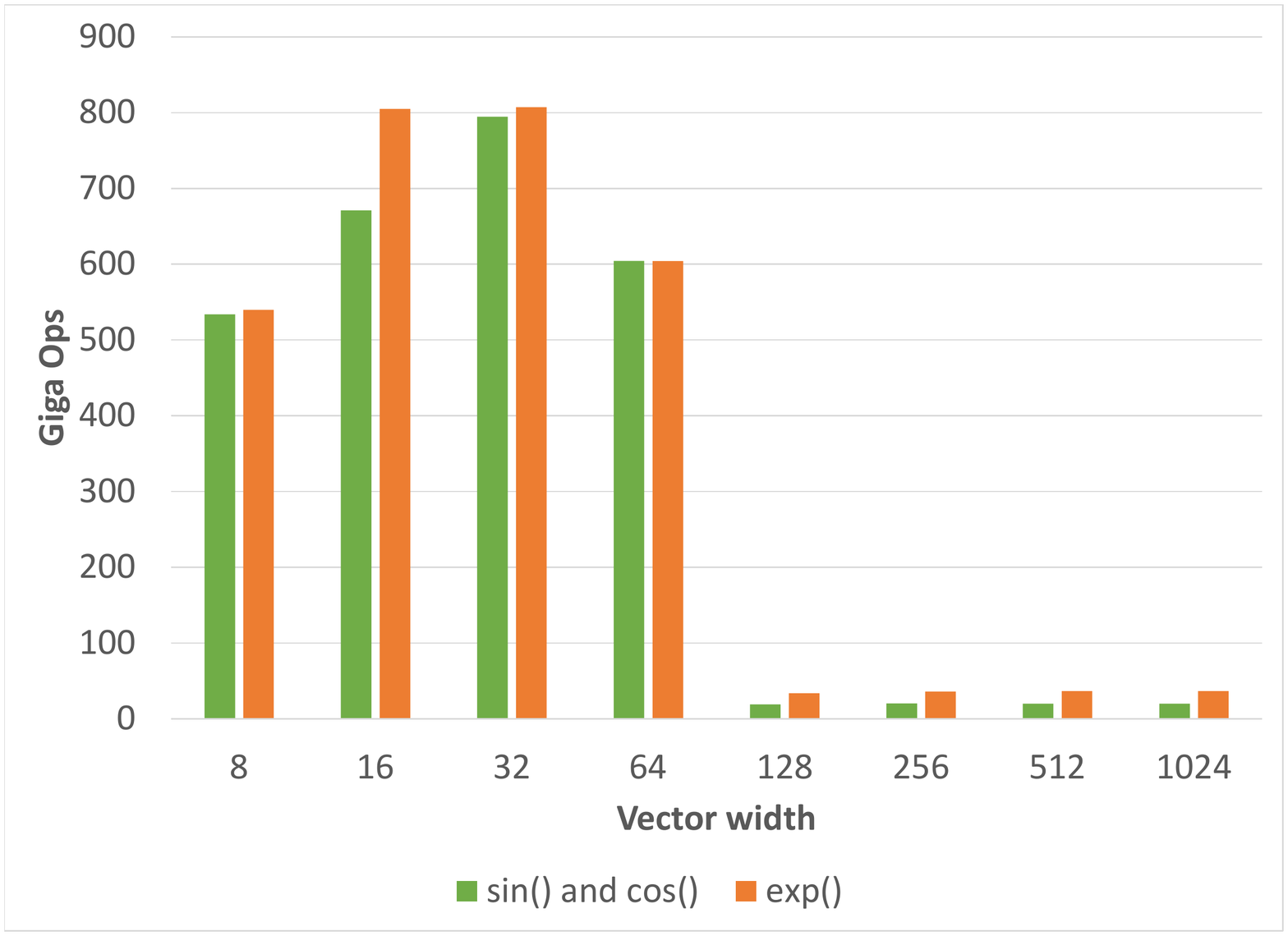}
\caption{The performance of KNL on a single compute node of
  Cori. Left: The floating point operations for computing a multiply and an add operations as functions of the length of the innermost vectorized loop. 
  Right: The performance of the chip for a sine and cosine operations per iteration, and one exponent operation per each loop iteration.}
\label{fig:flops}
\end{figure}

We first benchmarked the raw performance of KNL on a single compute node for double precision multiply/add floating point operations (left panel in Fig.~\ref{fig:flops}). Since, XFLAT extensively employs transcendental functions, we also benchmarked the performance of KNL for sin/cos, and exponent, respectively (right panel in Fig.~\ref{fig:flops}). In the innermost loop of the benchmark kernels simple floating point operations are performed on an array or vector of double-precision (DP) floating-point numbers. The widths of the vectors are chosen to be a multiple of that of the SIMD
registers of the computing component (512 bits or 8 DP for the Xeon Phi). We changed the length of the most inner loop from 8 (2\textsuperscript{3}) to 1024 (2\textsuperscript{10}). The same vector operations are repeated 10 million times in the middle loop to maintain data locality in the processor's cache. In the outermost loop all of the hardware threads are utilized to achieve the best performance.

As it is noticeable, the performance improved by increasing the vector width in the beginning, however, it dropped sharply by going beyond 64 in both plots. Part of the reason might be due to the internal caching mechanism that works well up to the size of 512 bytes. Similar results has been reported on KNC\cite{DBLP:journals/corr/NoormofidiAD15}. 

Next, we benchmarked XFLAT on a single KNL using two different memory mode. The MCDRAM on KNL can work on three different configurations: Cache mode, Flat mode, and Hybrid mode. In Cache mode MCDRAM acts as a cache layer between the processor's internal last level cache and DRAM. In Flat mode the MCDRAM acts as a software regular memory residing in the same address space as DRAM. In the Hybrid mode MCDRAM is the combination of the former two modes (part cache and part memory).

\begin{figure}[t]
\centering
\raisebox{7ex}{\includegraphics[width=.49\columnwidth]{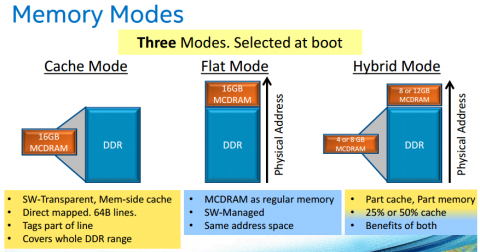}}
\includegraphics[width=.49\columnwidth]{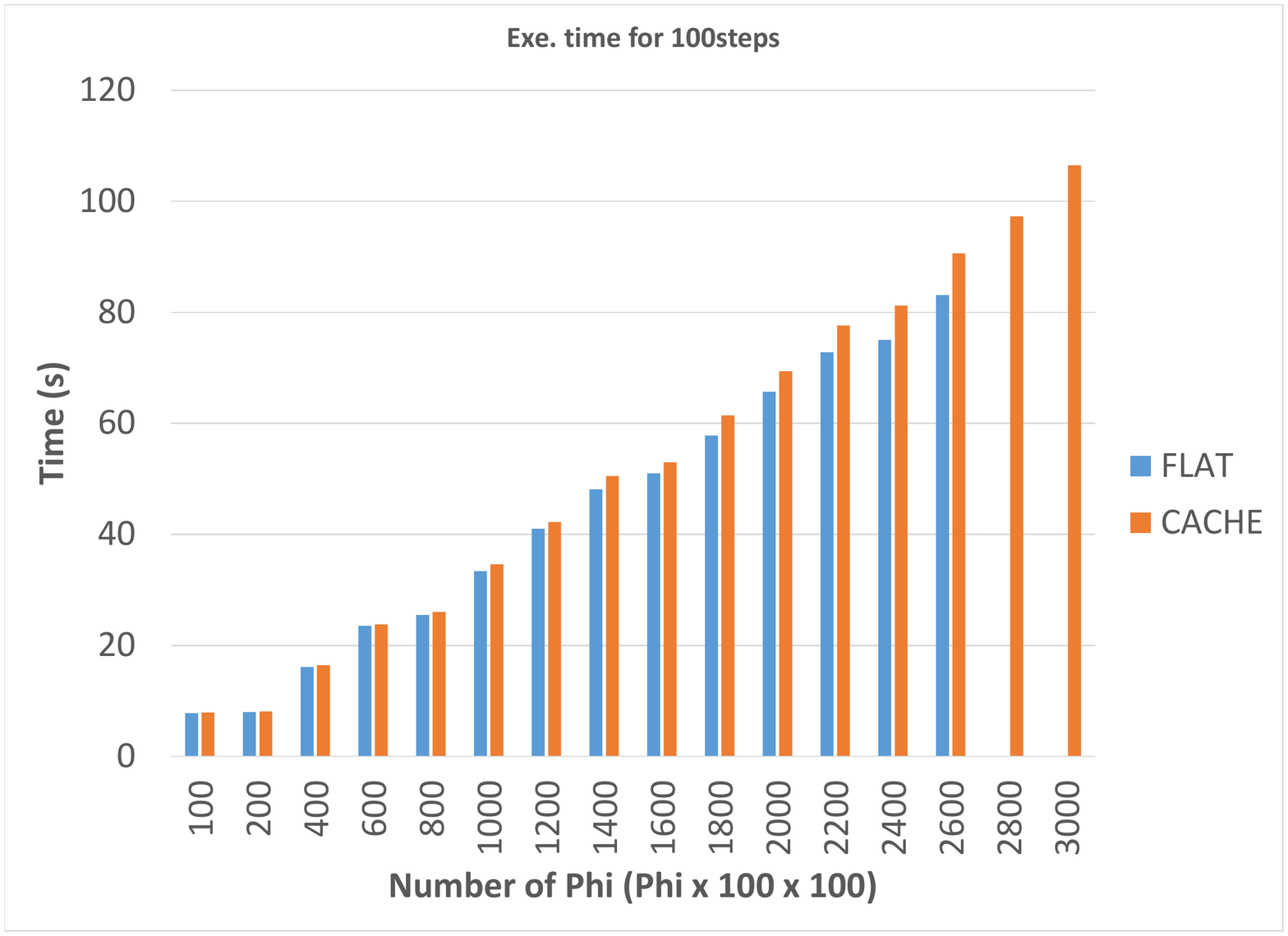}
\caption{Left: The KNL memory modes (image courtesy of NERSC).
  Right: The timing of XFLAT (calculating 100 radial steps) for the Flat and the Cache memory mode.}
\label{fig:FLAT-Cache}
\end{figure}

As shown in right panel of Fig.~\ref{fig:FLAT-Cache} the run times of these tests scale approximately linearly with problem size. Also the performance of Flat mode is slightly better, however, it requires that the application explicitly allocates memory on the second NUMA domain, since in the Flat mode the MCDRAM memory is shown as the second NUMA domain. In addition, the last two data points show that the memory footprint of XFLAT was too high (\textgreater 16 GB) to fit into MCDRAM in the Flat mode.

\subsection{Multi-node performance}

\begin{figure}
\centering
\includegraphics[width=.49\columnwidth]{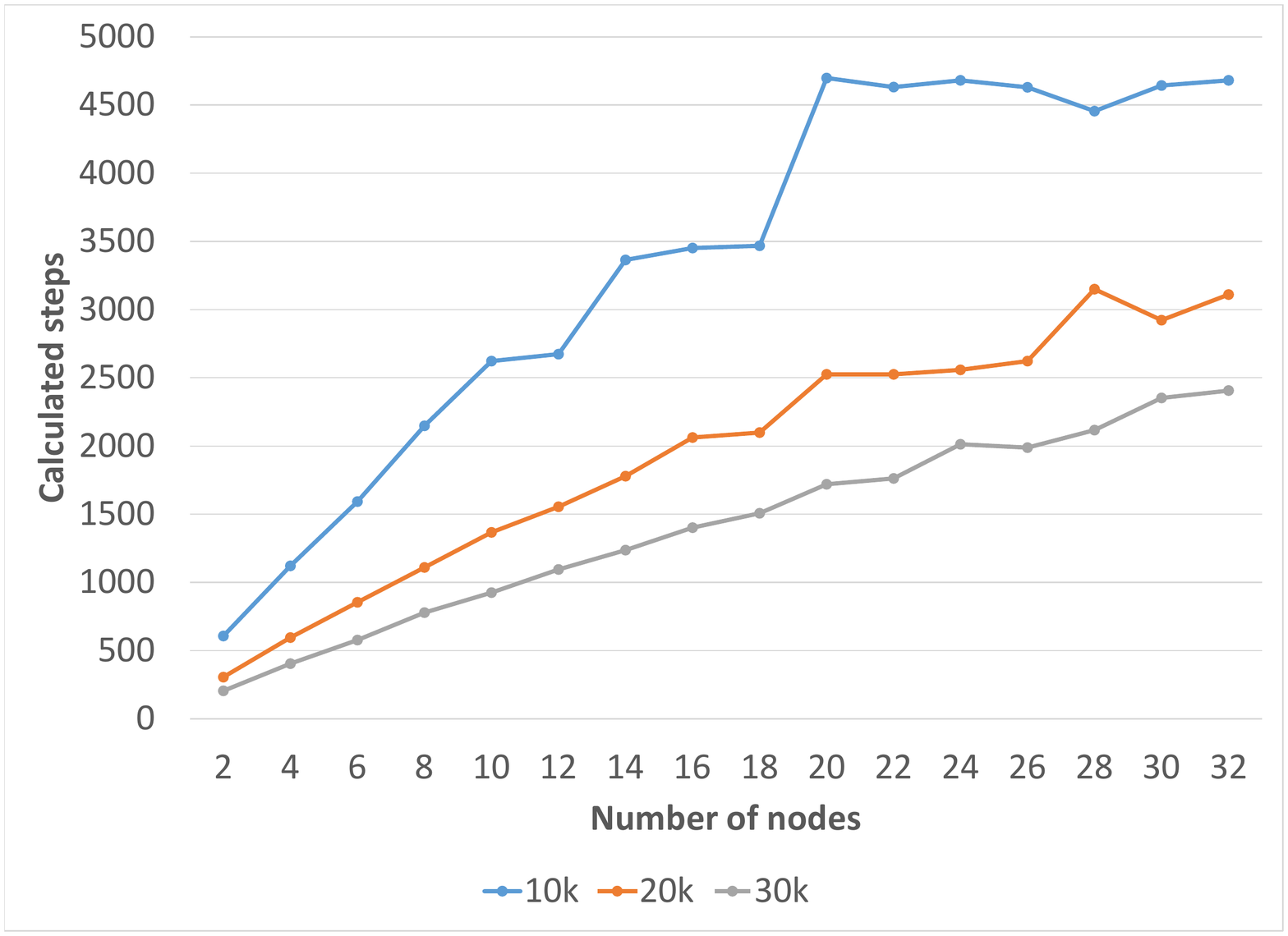}
\includegraphics[width=.49\columnwidth]{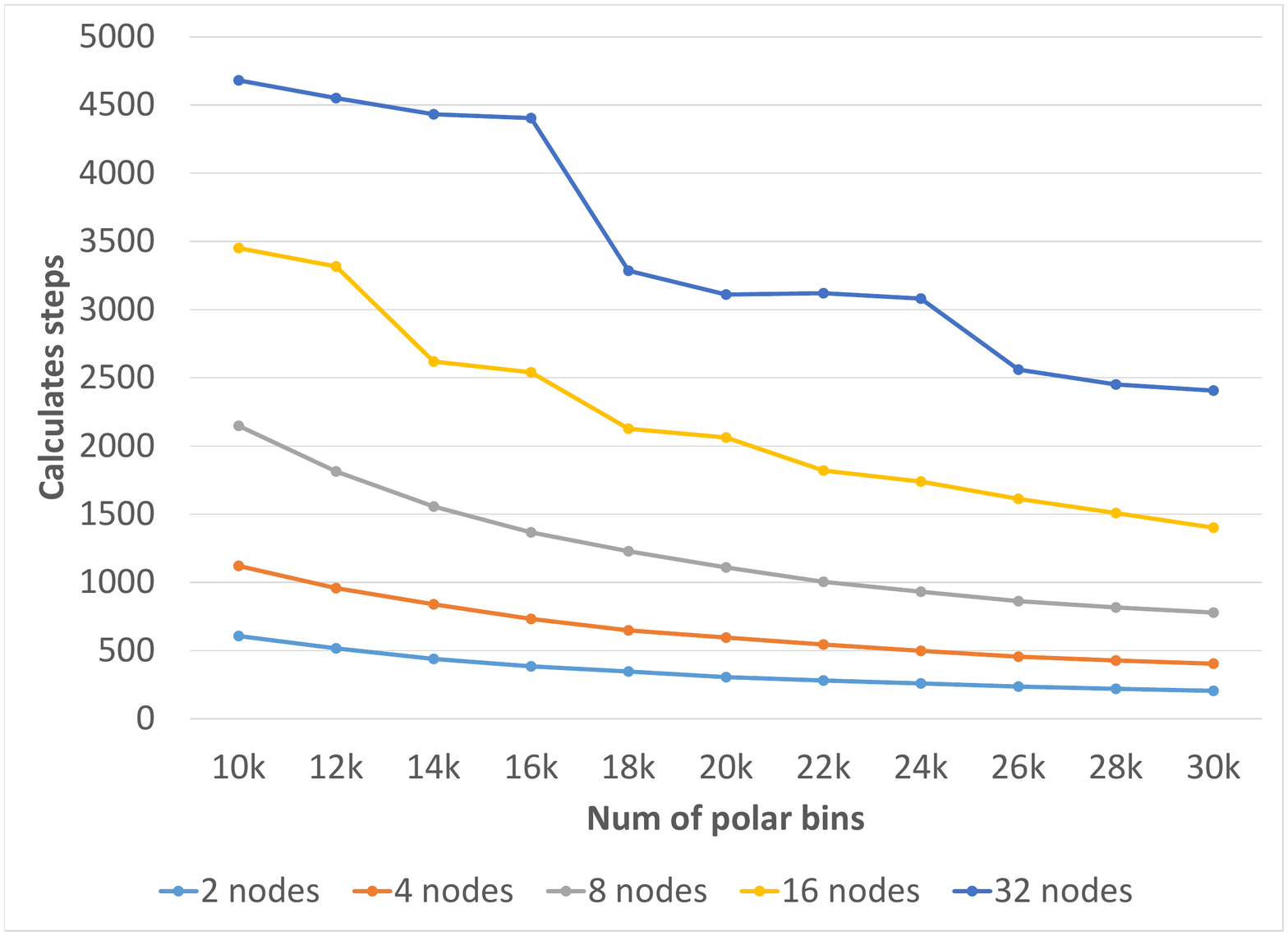}
\caption{Left: The number of calculated radial steps on computing 10k, 20k, and 30k angle bins in about 100 seconds. Right: The XFLAT performance on 2, 4, 8, 16, and 32 nodes with various number of polar bins (number of calculated radial steps per 100s).}
\label{fig:multi-node}
\end{figure}

We benchmarked XFLAT on multiple compute nodes. 
For the multi-node benchmarks we fixed the number of azimuth bins to 10 and number of energy bins to 100, then increased the number of polar angle bins in the ``standard problem'' from $10,000$ to $30,000$. The number of nodes were also varied from 2 to more than 32 nodes. In each of the tests we ran XFLAT without any disk I/O for approximately 100 seconds and computed the number of calculated steps per second. All the benchmarks were performed with MCDRAM configured in Cache mode. 
In the left panel of Fig.~\ref{fig:multi-node} the performance of XFLAT is shown as a function of number of compute nodes. The benchmarks were repeated for $10,000$, $20,000$, and $30,000$ polar angles. These benchmarks show 
that the performance of XFLAT scales well and if there are enough jobs (\textit{i.e.} polar angle bins), the scalibility is almost linear. The reason for the step pattern in the plot is due to having a load imbalance among threads/cores on KNL, as discussed in \cite{DBLP:journals/corr/NoormofidiAD15}. So the reason for the poor scaling behavior is because the 272 hardware threads of the Xeon Phi
cannot not be fully utilized for the studied problem size when many
compute nodes are used. For example, in the test with $10,000$ polar angles on 18 compute nodes each Xeon Phi process received almost
$555 (\approx10,000/18)$ polar angle bins, which is just a few more than twice the 272 threads available on the Xeon Phi. As a result, the OpenMP parallel \texttt{for} loop iterated three times in
each step with most of the threads idle in the last iteration. When
20 compute nodes were used, however, each Xeon Phi process received
500 polar angle bins, and the computation for these angle bins
was completed in two iterations by employing most of the threads. For low amount of load the performance of XFLAT does not increase much when more compute nodes are used because there is not enough load on each Xeon Phi (and consequently on each thread). 

In the right panel of Fig.~\ref{fig:multi-node} the performance of XFLAT is shown as a function of load (\textit{i.e.} polar angle bins) on various multi-node configurations. As one may notice again, distributing the load on many number of nodes will result in load imbalance on KNL's threads. 

\begin{figure}
\centering
\includegraphics[width=.49\columnwidth]{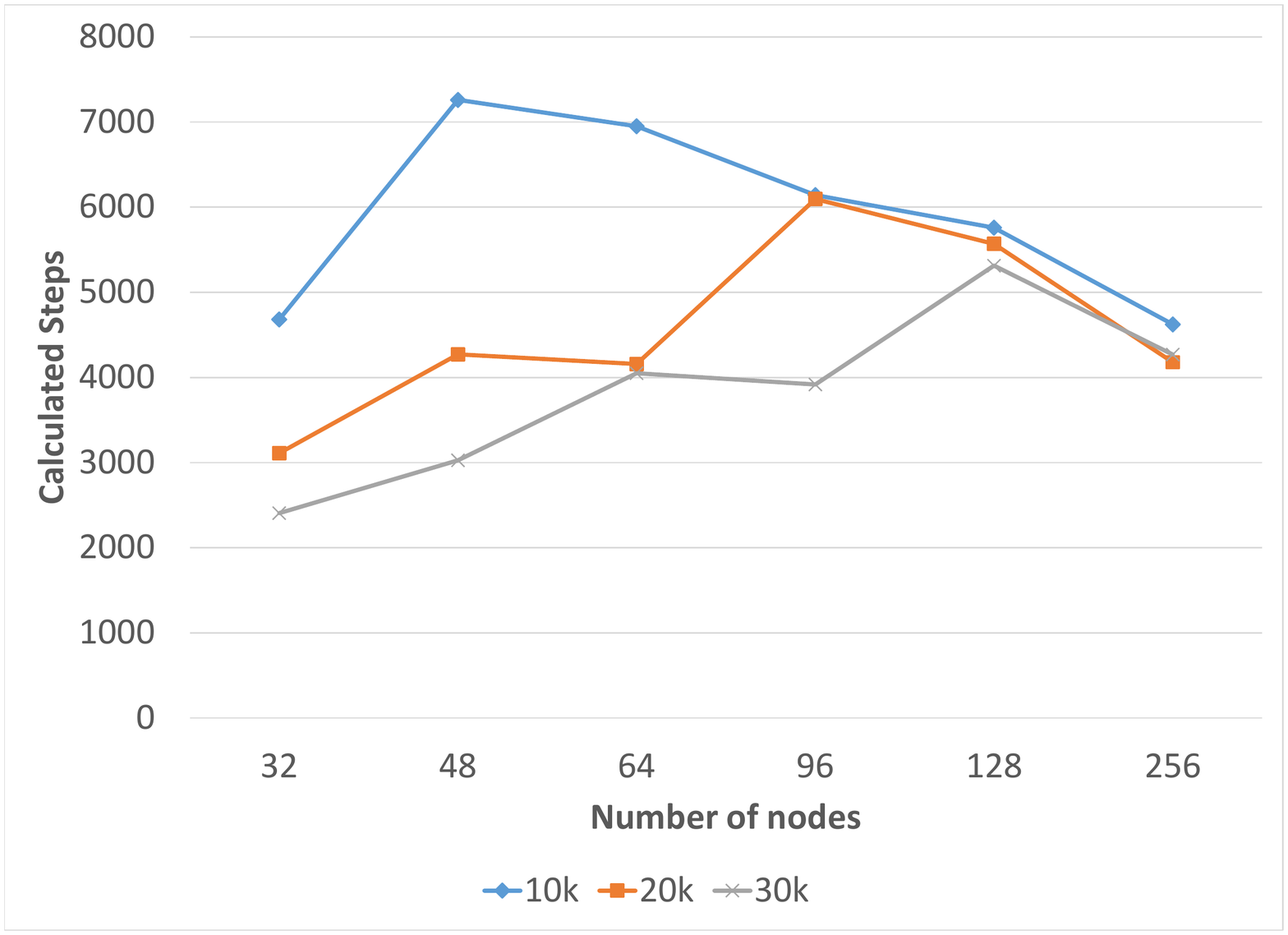}
\includegraphics[width=.49\columnwidth]{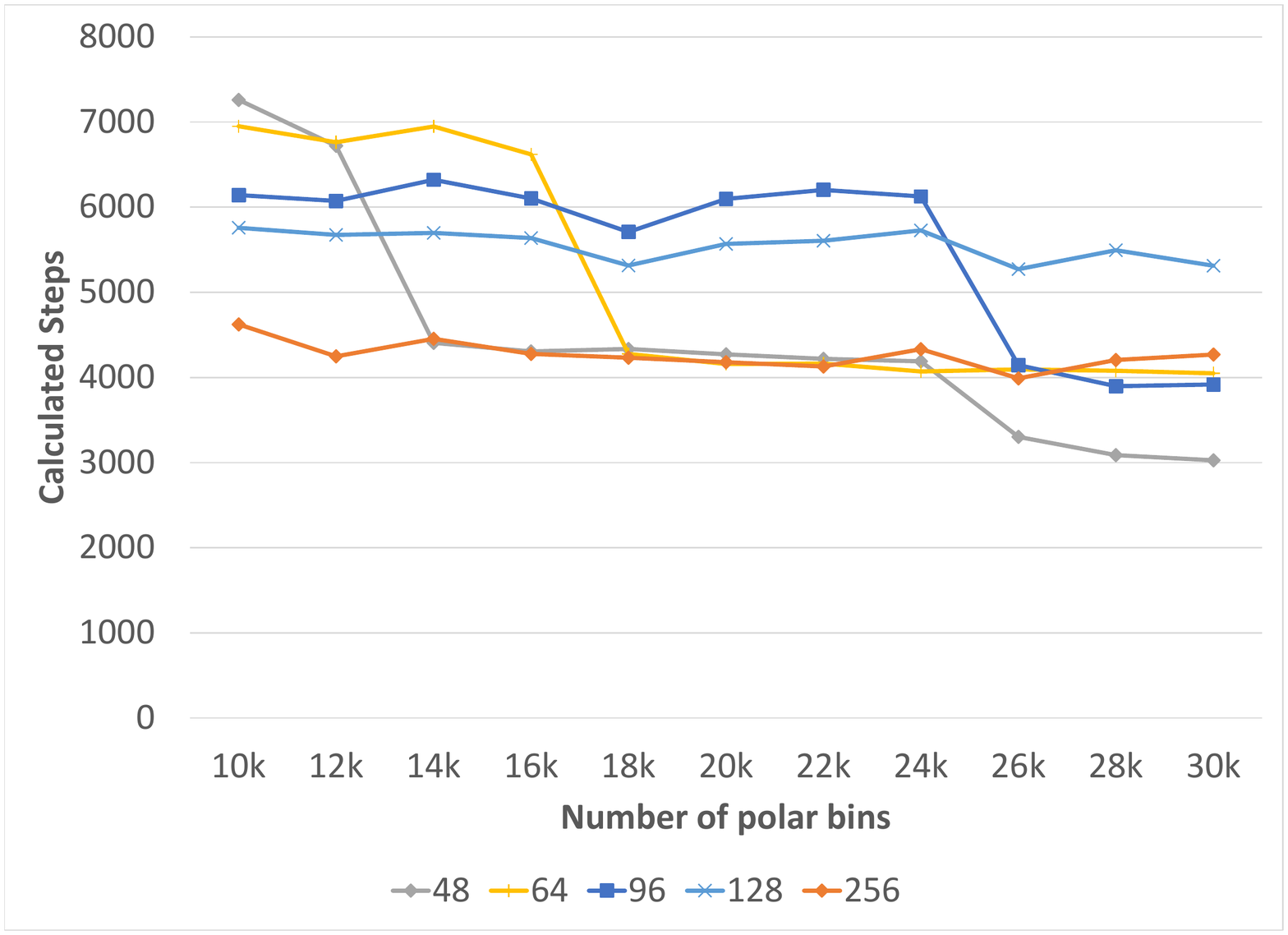}
\caption{Left: Scalibility of XFLAT on various number of nodes (2 to 256) for three different loads: 10k, 20k, and 30k polar bins. Right: XFLAT's performance when the number of nodes is fixed at 48, 64, 96, 128, or 256.}
\label{fig:sturate}
\end{figure}

Fig.~\ref{fig:sturate} shows the scalibility and saturation point of XFLAT. In the left panel, one can see the scalibility of the application for three different polar angle loads on up to 256 compute nodes. It is noticeable that by going to more compute nodes, the performance peaks and then drops, since there is not enough load (low number of polar bins). Adding each compute node result in increase the communications overhead to the system, however, individual threads do not have enough compute load to process. In the right panel, the number of nodes is fixed to between 48 and 256, but the load (number of polar bins) varies from 10k to 30k. For each node configuration, the performance is steady until reaching a critical point, then drops to the next level. This behavior is also due to the load imbalance on each KNL's threads which is caused by the change in the number of polar bins. The performance drop points for the run with 128 and 256 nodes only happen by going beyond 30k polar angle bins (not shown here).

\begin{figure}
\centering
\includegraphics[width=\textwidth,height=.5\textheight]{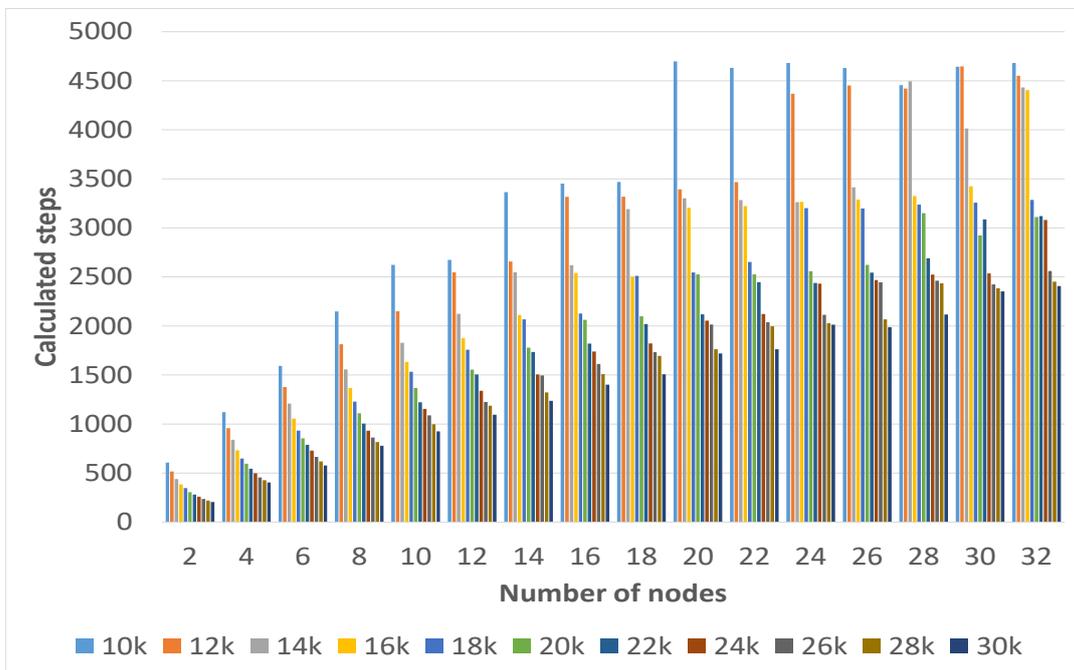}
\caption{The overall performance of XFLAT over various number of nodes and number of polar bins (calculated steps for 100s).}
\label{fig:multi-10-30}
\end{figure}

Fig.~\ref{fig:multi-10-30} depicts the full range of our benchmarks starting from $10,000$ polar bins to $30,000$ with $2,000$ incremental steps. It can be concluded that for a low amount of load, increasing the number of nodes does not result in performance improvement, since each hardware thread may not receive enough bins to keep the hardware fully loaded. For example, for the run with 10k polar bins, by going beyond 20 nodes the performance does not improve due to the same reason. On the contrary, the benchmark with 30k polar bins scale well even beyond 30 nodes. As a result, it is very important to choose the right number of compute node so that the load can distribute on all hardware thread evenly.

\section{Discussion and summary%
\label{sec:discussion}}

In this work we explore the Intel MIC architecture by
developing and benchmarking an astrophysical simulation code,
XFLAT, which compute neutrino oscillations in supernovae. Our
study is based on the extended (supernova neutrino) bulb model and
the algorithm outlined in Ref.~\cite{Duan:2008eb}. 
To fully utilize the SIMD units on the MIC architecture we have changed the
low-level module, i.e.\ the \NBeam{} module describing the neutrino
beam, from the AoS layout to 
the SoA layout. This approach boosts the performance on both the MIC and
CPU alike because modern CPUs also have SIMD units
albeit with smaller widths than those on the MIC. 
We have applied three levels of parallelization in XFLAT: MPI for
the inter-CPU/MIC parallelization, OpenMP for intra-CPU/MIC
parallelization, and SIMD for the most fine-grained parallelization
within each core.

We benchmarked
XFLAT on the Stampede supercomputer which is equipped with the
first-generation Xeon Phi or KNC. Our tests show that XFLAT achieves a
speedup of $2.5$--$2.9$ on a single Xeon Phi relative to a single
8-core E5 Xeon CPU, although the Xeon Phi on the Stampede has a peak
performance approximate 6 times  that of the CPU. 
We note that, however, the performance of the floating point
operations on the Xeon Phi depends on both the size of the
\texttt{omp} \texttt{for} \texttt{simd} loop and the nature of the
operations. For example, the performance of the double-precision 
\texttt{exp} function on the KNC is approximately 4 times of that on the
CPU \cite{Noormofidi:2015wia}.
The Xeon CPU has large caches per core than the KNC
which may have also play a role in our performance tests. 

\begin{figure}[t]
\centering
\includegraphics[width=.5\columnwidth]{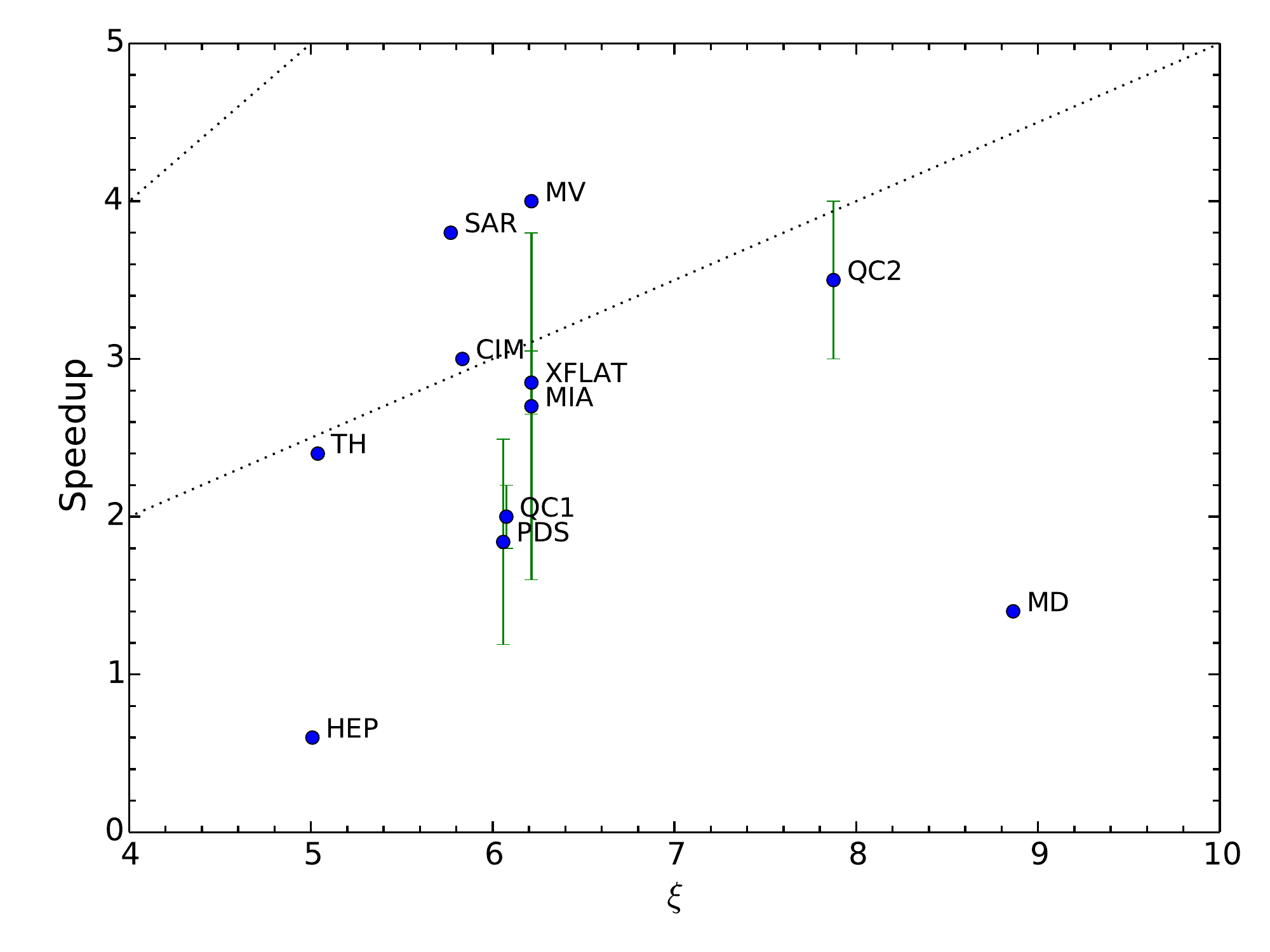}
\caption{The speedups because of the Xeon Phi coprocessor in various
  applications which were run on computer with various MIC-CPU peak
  performance ratios ($\xi$). The error bars
  represented the reported 
  the ranges of the speedups. The two dotted lines correspond to the
  scenarios with the speedup equal $\xi$ and $\xi/2$, respectively.}
\label{fig:speed-up}
\end{figure}

In Fig.~\ref{fig:speed-up} we compare the speedup of XFLAT on Stampede due
the Xeon Phis with those in applications for the
critical Ising model (CIM  \cite{Wende:2013:SMA:2503210.2503254}),
quantum chemistry (QC1 \cite{Apra:2014:EIM:2683593.2683667} and QC2
\cite{leang2014quantum}), 
thermo-hydrodynamics (TH \cite{Crimi2013551}),
experimental high-energy physics (HEP \cite{1748-0221-9-04-P04005}),
molecular visualization (MV \cite{Knoll:2013:RTV:2535571.2535594}),
molecular dynamics (MD \cite{pennycook2013exploring}),
dynamics of astrophysical objects (DAO \cite{Kulikov201571}), 
protein database search (PDS \cite{liu2014swaphi}),
synthetic aperture radar (SAR \cite{park2013efficient}) and
microscopic image analysis (MIA \cite{teodoro2014comparative}). In
presenting this comparison we plot the speedups because of the Xeon
Phi versus the peak performance ratio between the Xeon
Phi and CPU or
\begin{align}
\xi = \frac{P_\text{MIC}}{P_\text{CPU}},
\end{align}
where 
\begin{align}
P_\text{device} = (\text{width of the SIMD unit per core}) \times
(\text{number of cores per device}) \times
(\text{device clock rate}).
\end{align}
It seems that most of the
applications achieve a speed up of $\xi/2$ or less.

We have adopted the native mode of the Xeon Phi hoping that the same
code can run on both the CPU and the MIC.
However, because the I/O performance of the KNC is very poor compared
to the CPU, we have to implement different I/O modules for the MIC
and the CPU so that the CPU writes the data on behalf the Xeon Phi to the
disk. 

In multi-node tests we found that it was more difficult to fully utilize the
computing capability of the Xeon Phi than that of the CPU
because the Xeon Phi has
many more but less powerful cores than the CPU.
Further, it can also be a challenge to maintain the
load balance in a heterogeneous environment where both the CPU and
Xeon Phi are employed. 
There can be a significant drop in the performance of the Xeon Phi
when the number of jobs on the device is slightly more than a multiple
of the number of its hardware threads ($\sim 240$).

It has recently been shown that
the spherical symmetry and stationary assumption employed in the
extended bulb model 
could be broken spontaneously by neutrino oscillations
\cite{Duan:2014gfa,Abbar:2015fwa}. As a result, 
simulations in full 7-dimensional supernova models (with 1 temporal, 3
spatial and 3 momentum dimensions) must be performed in order to
study the real impacts of neutrino oscillations on 
supernova physics. This paradigm shift implies a dramatic increase of
several orders of 
magnitude in computational intensity which can be a good fit for the
next-generation MIC supercomputers. 

\section*{Acknowledgments}
This work was supported in part by DOE EPSCoR grant DE-SC0008142 (V.N.,
S.S.\ and H.D.) and
NSF grant OCI-1040530 (V.N.\ and S.R.A.) at UNM. We are
grateful to the Texas Advanced
Computing Center (TACC) and the UNM Center for
Advanced Research Computing (CARC) for providing computational resources
(Stampede and Bethe) used
in this work. We thank Dr.\ John Cherry and
Sajad Abbar for their assistance during the development of this
project.

\section*{References}
\bibliography{xflat}

\end{document}